\documentclass[11pt,a4paper]{article}
\usepackage[utf8]{inputenc}
\usepackage{a4wide}
\usepackage{amsmath,amssymb}
\usepackage{amsfonts}
\usepackage{cite}
\usepackage{color}
\usepackage{adjustbox}
\usepackage{multirow}
\usepackage{pdflscape}
\usepackage{textcomp}
\usepackage{gensymb}
\usepackage{graphicx}
\usepackage{subfigure}
\usepackage{diagbox}
\usepackage{pifont}
\usepackage{bigstrut}
\usepackage[small,bf]{caption}
\usepackage{array}
\usepackage{arydshln} 
\usepackage{booktabs}
\usepackage{stackengine}
\usepackage{mathtools}
\usepackage{enumerate}
\usepackage{makecell}
\usepackage{listings}
\usepackage{bbm}
\usepackage[unicode]{hyperref} 
\usepackage{authblk}

\definecolor{greenLinks}{rgb}{0, 0.6, 0} 
\definecolor{blueLinks}{rgb}{0, 0, 0.6}
\definecolor{redLinks}{rgb}{0.6, 0, 0}
\definecolor{eprintLinks}{rgb}{0.4, 0.4, 0.4}
\definecolor{journalLinks}{rgb}{0.6, 0, 0}
\DeclareMathOperator{\Tr}{Tr}
\DeclareMathOperator{\diag}{diag}

\makeatletter
\g@addto@macro\bfseries{\boldmath}
\makeatother

\allowdisplaybreaks
\numberwithin{equation}{section}

\begin{document}

\pagenumbering{Alph}
\begin{titlepage}

\begin{center}
{ \bf\LARGE {Dirac Neutrinos and Dark Matter within a Minimal \\[6pt] Discrete Symmetry Model }}\\[8mm]
Yakefu Reyimuaji$^{\,a,}$\footnote{E-mail: \href{mailto:yreyi@hotmail.com}{\texttt{yreyi@hotmail.com}}},
Murat Abdughani$^{\,a,}$\footnote{E-mail: \href{mailto:mulati@xju.edu.cn}{ \texttt{mulati@xju.edu.cn}}}

\vspace{8mm}
$^{a}$\,{\it School of Physical Science and Technology, Xinjiang University, Urumqi 830017, China} \\

\end{center}
\vspace{8mm}

\begin{abstract}
\noindent We present a model that extends the standard model by incorporating the simplest discrete symmetry groups, $Z_2$ and $Z_3$. This model introduces vector-like leptons and a real scalar singlet. Based on this framework, we generate Dirac neutrino masses and explain the neutrino normal mass ordering. The model also aligns well with current oscillation data regarding theoretical values of the leptonic mixing angles and the Dirac CP-violating phase. Furthermore, it predicts that the atmospheric mixing angle falls in the higher octant and proposes a viable dark matter candidate. We also discuss other phenomenological aspects and future testability of the model.

\end{abstract}

\end{titlepage}
\pagenumbering{arabic}
\setcounter{footnote}{0}


\section{Introduction}
\label{sec:intro}

Neutrino mass and dark matter represent significant drawbacks of the Standard Model (SM). Addressing either of these requires an extension of the SM through the construction of a new physics model. The experimental discovery of neutrino oscillations serves as a clear indicator of non-zero neutrino mass and mixing~\cite{McDonald:2016ixn,Kajita:2016cak}. However, several aspects remain unresolved, including the nature of the neutrino (Dirac or Majorana), mass ordering, precise values of some mixing angles, and the leptonic CP-violating phase. Similarly, while there is overwhelming observational evidence for the existence of dark matter (DM)~\cite{Bertone:2004pz,Garrett:2010hd,Cirelli:2024ssz}, questions about its particle nature and production mechanisms, such as thermal freeze-out~\cite{Zeldovich:1965gev,Scherrer:1985zt,Steigman:2012nb,Frumkin:2022ror} or freeze-in~\cite{McDonald:2001vt,Hall:2009bx,Bernal:2017kxu}, persist. Neutrinos and DM share characteristics as weakly interacting, invisible, and elusive particles, making it compelling to unify them within a single theoretical framework and explore potential dual-purpose solutions.

The distinction between Dirac and Majorana neutrinos leads to two different mass generation mechanisms. Neutrinos are often expected to be Majorana fermions, primarily because they are color and charge neutral, and the model-independent effective field theory approach—namely, the seesaw mechanism—most naturally and elegantly generates Majorana masses. However, the absence of experimental confirmation from neutrinoless double beta decay experiments, despite considerable efforts, suggests it is premature to dismiss the possibility of Dirac neutrinos. Recent studies have shown that the seesaw mechanism can be adapted to generate Dirac neutrino masses at the tree level, referred to as the Dirac seesaw~\cite{Gu:2006dc,Valle:2016kyz,Reig:2016ewy,Bonilla:2016zef,Ma:2016mwh,Wang:2016lve,Yao:2018ekp,CentellesChulia:2018bkz}. An intriguing feature of this mechanism is that it also simultaneously addresses other beyond-SM issues, such as DM~\cite{deGouvea:2015pea,Ma:2015mjd,CentellesChulia:2016rms,CentellesChulia:2016fxr,Borah:2016zbd,Borah:2016hqn,Borah:2017leo,CentellesChulia:2017koy,CentellesChulia:2018gwr,CentellesChulia:2018bkz,Borah:2018gjk,Borah:2019bdi,Gu:2019ohx,Gu:2019ird,Nanda:2019nqy,De:2021crr,Borah:2022obi,Singh:2024imk,Borah:2024gql}, the strong CP problem~\cite{Baek:2019wdn,Peinado:2019mrn,delaVega:2020jcp,Dias:2020kbj,Berbig:2022pye,Penedo:2022gej}, and leptogenesis~\cite{Dick:1999je,Murayama:2002je,Cerdeno:2006ha,Gu:2016hxh,Narendra:2017uxl,Gu:2019yvw}. The recent surge of interest in the dynamical generation of the Dirac neutrino mass, alongside simultaneous solutions for other outstanding challenges to the SM, highlights critical features testable by experiments.

Given their feebly-interacting properties, the underlying theories of DM and neutrino mass generation may be interconnected. In certain models, neutrinos play a role analogous to DM, although they are generally considered distinct. In particular, some models posit DM as essential for mediating interactions necessary for neutrino mass generation, such as in scotogenic~\cite{Ma:2006km,Hirsch:2013ola,Borah:2016lrl}, dark linear~\cite{CarcamoHernandez:2023atk,Batra:2023bqj}, and inverse seesaw mechanisms~\cite{Mandal:2019oth,CentellesChulia:2020dfh}. Moreover, in Dirac seesaw mechanisms, the symmetry envisioned to preserve the lepton number and prohibit Majorana mass terms also stabilizes the DM~\cite{deGouvea:2015pea,Ma:2015mjd,CentellesChulia:2016rms,CentellesChulia:2016fxr,CentellesChulia:2017koy,CentellesChulia:2018gwr,CentellesChulia:2018bkz,Gu:2019ohx,Gu:2019ird}. Along this line of research, model building or a specific realization of the existing model in which the Dirac neutrino mass is generated by the seesaw mechanism with heavy fermionic DM is of great scientific importance~\cite{Ma:2021szi,Ma:2021bzl}. In addition, the classification of the inverse seesaw family for Dirac neutrinos~\cite{CentellesChulia:2020dfh} and the specific realization with DM~\cite{Gu:2019gzy} is explored.

The $Z_3$ symmetry is the minimal symmetry that forbids the right-handed neutrino Majorana mass terms, thus preserving their Dirac nature. This symmetry could originate from spontaneous breaking of larger symmetries, such as gauged $U(1)_L$~\cite{Ma:2021fre}, $U_{B-L}$~\cite{Ma:2015mjd}, or emerge through lepton flavor triality in $A_4$ symmetry~\cite{Ma:2010gs,He:2006dk,Cao:2011df}. In pursuit of such a minimal framework, we present a concrete realization of the Dirac seesaw mechanism based on $Z_2\times Z_3$ symmetry that simultaneously explains neutrino mass generation, flavor mixing, and the observed DM relic abundance. 

The paper is organized as follows: section~\ref{sec:model} presents the model construction, specifying symmetry assignments, field content, and the Yukawa Lagrangian governing lepton masses. We further analyze the scalar potential, which determines the mass spectrum and interactions of the scalar sector. Section~\ref{sec:phnmlgy} details the phenomenological consequences, demonstrating that the model successfully accommodates neutrino oscillation data with normal mass ordering and explains the consistent DM relic density. We additionally examine constraints from lepton flavor violation processes and LHC searches for exotic leptons. Section~\ref{sec:conclusions} summarizes our findings and discusses their broader implications. The technical aspects of the analysis are provided in two appendices.

\section{The model}
\label{sec:model}
It is well-known that the SM, being a low-energy effective theory, is incomplete and must be augmented at higher energies. In what follows, we try to construct such a UV theory based on guiding principles of symmetry. In adherence to the symmetry principle and the seesaw idea, our main goal in this section is to forbid the usual Majorana mass term for the right-handed neutrino, while achieving the Dirac mass term $m_D\overline{\nu_L}\nu_R$ after integrating out the heavy states. To this end, we introduce a model extending the SM by incorporating $Z_2\times Z_3$ symmetries and by adding three generations of right-handed neutrinos $\nu_R$ and vector-like lepton doublets $\psi_{L,R}= (\psi^0_{L,R},\psi^-_{L,R})^T$. Additionally, the scalar sector includes a real singlet $\sigma$, besides the Higgs doublet $H$. The electroweak gauge quantum numbers and the transformation properties under the discrete symmetries are specified in Table~\ref{tab:chasign}. As indicated, all leptons transform similarly under the $Z_3$ symmetry, while the scalars are invariant. For the $Z_2$ symmetry, the SM fields are invariant, whereas the newly introduced fields change sign.

\begin{table}[h]
	\begin{center}
		\begin{tabular}{|p{5em} p{4em} p{4em}  p{4em} p{4em} p{4em} | p{3em} p{1em}|}
			\hline 
			Fields       & $L_i$   &  $e_{i,R}$   &  $\nu_{i,R}$         & 
			$\psi_{i,L}$     &  $\psi_{i,R}$  &  $H$             &  $\sigma$                            \\
			\hline 
			$SU(2)_\mathrm{L}$   & $2$   & $1$   & $1$        &  
			$2$         & $2$         &  $2$             & $1$                           \\
			$U(1)_\mathrm{Y}$    & $-\frac{1}{2}$   & $-1$    &  $0$     &
			$-\frac{1}{2}$       & $-\frac{1}{2}$       &  $\frac{1}{2}$  & $0$                       \\
			$Z_3$      & $\omega$     &  $\omega$     &  $\omega$        &   
			$ \omega$  & $\omega$                       &  $1$     &         $1$            \\
			$Z_2$      & $+1$     &  $+1$     &  $-1$        &   
			$-1$  & $-1$                  &  $1$     &         $-1$            \\
			\hline
		\end{tabular}
	\end{center}
	\caption{ The electroweak $SU(2)_\mathrm{L}\times U(1)_\mathrm{Y}$ gauge quantum numbers and $Z_2$, $Z_3$ transformation properties of the left-handed lepton doublets $L_i$ (with $i=1,2,3$), the right-handed neutrinos $\nu_{i,R}$, the vector-like leptons $\psi_{i,L}$ and $\psi_{i,R}$, the Higgs doublet $H$ and the scalar singlet $\sigma$. Here, $\omega$ is a cubic root of unity, e.g. $\omega = e^{2\pi i/3}$.}
	\label{tab:chasign}
\end{table}

\begin{figure}[!ht]
	\centering
	\includegraphics[width=0.5\linewidth]{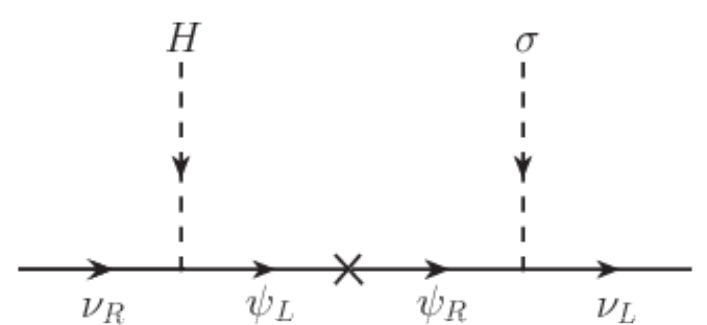}
	\caption{The  tree level Feynmann diagram for neutrino mass generation.}
	\label{fig:Drcssfynm}
\end{figure}

The Yukawa interactions between the scalars and leptons are expressed in the following Lagrangian:
\begin{equation}
	-\mathcal{L}_Y = Y_{E,ij}\bar{L}_i H e_{j,R}+  Y_{\psi,ij}\bar{L}_i  \psi_{j,R} \sigma +  Y_{\nu,ij}\overline{\psi_{i, L}} \tilde{H} \nu_{j,R} + \mathrm{h.c.},
    \label{eq:yukawalag}
\end{equation}
where $\tilde{H} = i \sigma^2 H^*$. Although the Yukawa interaction $\bar{L}\tilde{H}\nu_{R}$, which is needed for the neutrino Dirac mass term, is allowed by $Z_3$ symmetry, it is forbidden by the $Z_2$. This mass term is generated when the latter is spontaneously broken by vacuum expectation value (vev) of $\sigma$ and after the heavy vector-like leptons are integrate out. Consequently, the small neutrino mass is explained. The $Z_3$ symmetry disallows the Majorana mass term for the right-handed neutrino, even after the scalars gain vevs, thus preserving the Dirac nature of the neutrino. The mass term for vector-like leptons remains invariant under all symmetries and reads $M_{ij}\overline{\psi_{i, L}}\psi_{j, R}$. This mass term is expected to be large, which suppresses the neutrino mass via the seesaw mechanism, depicted in Figure~\ref{fig:Drcssfynm}.

The scalar potential, invariant under both SM and the extended symmetries $Z_2 \times Z_3$, is given by
\begin{equation}
	V(H,\sigma)= -\mu_{H}^2 H^\dagger H -\frac{1}{2}\mu_{\sigma}^2 \sigma^2 + \frac{\lambda_H}{2} ( H^\dagger H )^2  + \frac{\lambda}{2} ( H^\dagger H )\sigma^2+\frac{\lambda_\sigma}{4}\sigma^4 .
	\label{eq:sclrptn}
\end{equation}
All couplings in this potential can be made real. The potential being bounded from below requires that the quartic couplings satisfy $\lambda_H \ge 0, \lambda_\sigma \ge 0$ and $\lambda +\sqrt{2\lambda_H \lambda_\sigma} \ge 0$. Expanding the scalar fields around their vevs, $H=(G^+, (v+H^0+iG^0)/\sqrt{2})^T$, $\sigma=(u+\sigma^\prime)/\sqrt{2}$, we obtain stationary conditions for the potential:
\begin{equation}
	\begin{aligned}
		\mu^2_H=& \frac{1}{4}\left(2\lambda_H v^2+\lambda u^2 \right),  \\
		\mu^2_\sigma=& \frac{1}{2}\left(\lambda v^2+\lambda_\sigma u^2 \right).
	\end{aligned}
\end{equation}
Given the expansions of the scalar expansions, we rewrite the potential in terms of the neutral and charged scalar components and subsequently extract the mass matrix for the two neutral scalars,
\begin{equation}
	M^2_{H^0,\sigma^\prime}= 
	\begin{pmatrix}
		\lambda_H v^2 & \frac{1}{2} \lambda uv \\ \\[0.1pt]
		\frac{1}{2}  \lambda uv & \frac{1}{2}\lambda_\sigma u^2
	\end{pmatrix} .
\end{equation}
This mass matrix is diagonalized by a $2 \times 2$ orthogonal rotation matrix characterized by an angle $\theta$, which satisfies the relation $ \tan2\theta= \frac{2\lambda uv }{\lambda_\sigma u^2-2\lambda_H v^2}$. Diagonalization yields two physical states, $h$ and $\phi$, with the respective masses given by
\begin{equation}
	\begin{aligned}\label{eq:sclrmas}
		m^2_h =& \frac{1}{4} \left[2\lambda_H v^2  + \lambda_\sigma u^2 - \sqrt{\left(2\lambda_H v^2  - \lambda_\sigma u^2\right)^2+(2\lambda u v)^2}\, \right], \\
		m^2_\phi =& \frac{1}{4} \left[2\lambda_H v^2  + \lambda_\sigma u^2 + \sqrt{\left(2\lambda_H v^2  - \lambda_\sigma u^2\right)^2+(2\lambda u v)^2}\, \right].
	\end{aligned}
\end{equation}
The transformations between the original scalar fields and the mass eigenstates are governed by the following linear relationships:

\begin{equation}
	\begin{aligned}
		h=& \cos \theta H^0 + \sin \theta \sigma^\prime, \\
		\phi = &\cos \theta \sigma^\prime -\sin \theta H^0.
	\end{aligned}
\end{equation}
We identify $h$ as the SM-like Higgs boson with a mass of $m_h = 125$ GeV, and $\phi$ as another neutral scalar, which is heavier than the Higgs. 

The expression for the mixing angle $\theta$ and the scalar potential parameter $\lambda$, as shown in eq.~\eqref{eq:sclrptn}, are crucial in determining the degree of mixing between these two scalars. Notably, in the absence of $\lambda$, there is no mixing ($\theta = 0$), leading the scalars to decouple, as anticipated. On the other hand, the same parameter determines whether or not these scalar masses degenerate. Specifically, if $\lambda =0$, there exists a parameter space where the relation for two quartic couplings $\lambda_\sigma =2(v^2/ u^2)\lambda_H$ holds, resulting in equal mass squares, $m^2_h=m^2_\phi=\lambda_H v^2$. Conversely, when $\lambda$ is nonzero, it is not possible to achieve mass equality. If $\lambda \neq 0$, the minimum mass-squared difference, $m^2_\phi-m^2_h=\lambda u v$, occurs under the condition that the aforementioned relationship between $\lambda_\sigma$ and $\lambda_H$ is maintained. However, mass degeneracy is prohibited in this scenario. Additionally, the vacuum expectation values $u$ (also $v$) influence the masses of these scalars. For instance, if  $v\gg u=0$, $m_h$ vanishes, and  resulting in $m^2_\phi =\lambda_H v^2> m^2_h$. Conversely, when $u\gg v$, the mass squared of $h$ becomes $m^2_h\simeq (\lambda_H -\frac{\lambda^2}{2\lambda_\sigma})v^2+\mathcal{O}(v^2/u^2)$, which is significantly smaller than $m^2_\phi \simeq \frac{1}{2}(\lambda_\sigma u^2+\frac{\lambda^2}{\lambda_\sigma}v^2)+\mathcal{O}(v^2/u^2) $, assuming all quartic couplings are of similar magnitude. Therefore, one can qualitatively expect the mass of  $\phi$ to be substantially larger than that of the Higgs boson without detailed analysis of these couplings. Given the definition of the mass basis, the alternative assignment of the heavier scalar as the SM Higgs and the lighter one as the new neutral scalar is also valid. A similar discussion can be applied to this configuration.
Additionally, other would-be Nambu-Goldstone bosons, i.e., charged and pseudo-scalar components of the Higgs doublet, remain massless and become the longitudinal components of the $W$ and $Z$ bosons.

Regarding the fermionic sector of the model, masses are generated after the spontaneous breaking of electroweak and extended symmetries. The charged lepton mass matrix, in the bases of $\left(\overline{e_{i,L}}, \overline{\psi^-_{i,L}}\right)$ and $(e_{j,R}, \psi^-_{j,R})^T$, is  given by
\begin{equation}\label{eq:MEmatrM}
	M_E= \begin{pmatrix}
		m_E &  m_\psi \\
		0     &  M
	\end{pmatrix},
\end{equation}
where each entry consists of $3\times 3$ block matrices $m_E \equiv \frac{1}{\sqrt{2}}Y_E v$, $m_\psi \equiv \frac{1}{\sqrt{2}}Y_\psi u$ and $M$. There is a mixing between the left-handed charged lepton and the right-handed component of the vector-like lepton, while no mixing occurs between the left-handed vector-like lepton and the right-handed charged lepton due to the $Z_2$ symmetry. It can be diagonalized by
\begin{equation}\label{eq:flldiagM}
	U_L^\dagger M_E V_R = \diag(m_i, M_i).
\end{equation} 
Details of the diaonalziation are provided in Appendix~\ref{app:Dialepmas}.

In the neutral lepton sector, the mass matrix in the bases of $\left(\overline{\nu_{i,L}}, \overline{\psi^0_{i,L}}\right)$ and $(\nu_{j,R}, \psi^0_{j,R})^T$ is given by
\begin{equation}\label{eq:NumatrM}
	M_\nu= \begin{pmatrix}
		0 &  m_\psi \\
		m_D    &  M
	\end{pmatrix},
\end{equation}
where the Dirac mass matrix, $m_D$, is defined as $m_D \equiv \frac{1}{\sqrt{2}}Y_\nu v $, mirroring the structure observed in the charged lepton sector. Each block of the neutral lepton mass matrix consists of $3\times 3$ submatrices. The block diagonalization with seesaw idea yields the light neutrino mass matrix
\begin{equation}\label{eq:lghtnumasM}
	m_\nu=-m_\psi M^{-1}m_D= - \frac{uv}{2}Y_\psi M^{-1} Y_\nu .
\end{equation}
The masses of the light neutrinos and the neutral components of the vector-like leptons are achieved by full diagonalization of the mass matrix in eq.~\eqref{eq:NumatrM}
\begin{equation}
\begin{aligned}
	&\mathcal{U}_L^\dagger M_\nu \mathcal{V}_R = \diag \left(m^{(\nu)}_i, M^{(0)}_i\right).
\end{aligned}
\end{equation}

Notice that this approach highlights the mechanism by which the small Dirac neutrino masses arise, featuring double suppression through the large mass scale $M$ and small Yukawa couplings $Y_\psi$.

\section{Phenomenological analysis}
\label{sec:phnmlgy}

The previous section detailed the origins of Dirac neutrino mass, explaining its smallness through a vector-like lepton mediated seesaw mechanism. This section focuses on the phenomenological consequences, encompassing concrete realizations of neutrino masses, mixing, leptonic CP violation, and the explanation of DM.

\subsection{Neutrino masses and mixing}
\label{subsec:neuphnmlgy}
Neutrino masses are derived from the mass matrix in eq.~\eqref{eq:lghtnumas} through the diagonalization procedure outlined in eq.~\eqref{eq:numsdiag}. Before settling on a basis, it is evident that there are three basis-independent constraints between the neutrino mass matrix elements and the squares of the neutrino masses:
\begin{equation}
		\Tr (m_\nu m_\nu^\dagger )  = \sum_{i,j=1}^3 \left| m_{\nu,ij}\right|^2 = \sum_{i=1}^3 \left[m^{(\nu)}_i\right]^2, 
		\label{eq:trrelat}
\end{equation}
\begin{equation}
		\det (m_\nu m_\nu^\dagger ) = \left|\det (m_\nu) \right|^2 = \prod_{i=1}^{3} \left[m^{(\nu)}_i\right]^2, 
		\label{eq:dterminv}
\end{equation}
\begin{equation}
	\begin{aligned}\label{eq:maynrin}
		\frac{1}{2}\left(\Tr[m_\nu m_\nu^\dagger ]^2 - \Tr[(m_\nu m_\nu^\dagger )^2]\right)= & \frac{1}{2}\sum \varepsilon_{ijk}^2 \varepsilon_{abc}^2 \left|m_{\nu,jb}\right|^2 \left|m_{\nu,kc}\right|^2 \\
		& -\sum \varepsilon_{ijk}^2 \mathrm{Re} \left(2m_{\nu,ik} m_{\nu,jj}m_{\nu,ij}^*m_{\nu,jk}^*+m_{\nu,ii} m_{\nu,jj}m_{\nu,ij}^*m_{\nu,ji}^*\right)\\
		&=\left[m^{(\nu)}_1 m^{(\nu)}_2\right]^2+\left[m^{(\nu)}_1 m^{(\nu)}_3\right]^2+\left[m^{(\nu)}_2 m^{(\nu)}_3\right]^2,
	\end{aligned}
\end{equation}
where all summed indices range from 1 to 3. These equations not only remain independent of the basis choice but also provide connections between particular combinations of mass matrix elements on the left and functions of the physical neutrino masses on the right. Specifically, eqs.~\eqref{eq:trrelat}, ~\eqref{eq:dterminv} and ~\eqref{eq:maynrin} are the linear, cubic, and quadratic invariants of the neutrino mass squares, respectively. For instance, eq.~\eqref{eq:trrelat} indicates that the sum of the modulus squares of the mass matrix elements should not exceed the sum of the squares of the three light neutrinos' masses. According to recent global fit results for neutrino oscillation~\cite{deSalas:2020pgw,10.5281/zenodo.4726908,Capozzi:2021fjo,Gonzalez-Garcia:2021dve}, even the largest modulus square of the mass matrix cannot exceed  $\mathcal{O}(10^{-3})$ eV$^2$, assuming the lightest neutrino mass is zero. 

While this provides a qualitative overview of the small neutrino mass generation, a similar flavor issue as in the SM persists. The relative sizes of these matrix elements are not inherently constrained by the theory; rather, they are restricted by experimental results. Consequently, all permissible elements can vary widely, leaving the observed hierarchy in lepton masses and the form of their mixing matrix unexplained. This issue could be addressed by introducing a flavor symmetry, which would shape the forms of mass matrices. For illustrative purposes, one might consider the implementation of models such as the $A_4$ flavor symmetry~\cite{Altarelli:2005yp,Memenga:2013vc,Borah:2017dmk,Borah:2018nvu,Ding:2020vud}, or explore scenarios involving extra-dimensional warped geometries~\cite{Chen:2015jta}. For instance, implementing an $A_4$ flavor symmetry, where the SM lepton representations are $L\sim~ 3$, $e_{1,R}\sim 1$, $e_{2, R}\sim 1'^\prime$, $e_{3,R}\sim 1^\prime$, and a flavon triplet with uniform vacuum expectation values in all field directions, constrains the charged lepton mass matrix to the following form:\footnote{We choose to work in $A_4$ group generator $S$ diagonal basis of the triplet representation.}
\begin{equation}
	m_{E} = \frac{v}{\sqrt{2}} \begin{pmatrix}
		y_e & y_\mu & y_\tau \\
		y_e & \omega^2 y_\mu & \omega y_\tau  \\
		y_e & \omega y_\mu & \omega^2 y_\tau
	\end{pmatrix}.  
\end{equation}
This matrix can be diagonalized, $U_{L}^{(e)\dagger} m_Em_E^\dagger U_{L}^{(e)} = \diag(m_{e}^2, m_{\mu}^2,m_{\tau}^2)$, by the unitary matrix
\begin{equation}
	U_{L}^{(e)} = \frac{1}{\sqrt{3}} \begin{pmatrix}
			1 & 1 & 1\\
			1 & \omega & \omega^2 \\
			1 & \omega^2 & \omega
		\end{pmatrix},
\end{equation}	
yielding charged lepton masses $m_\alpha = \sqrt{3/2} \left|y_\alpha \right|v$ for $\alpha = e, \mu, \tau$. Such symmetry-driven constraints not only provide a theoretical basis for the mass hierarchy and mixing patterns observed experimentally but also offer a structured pathway for future theoretical and experimental investigations.

In expanding our discussion on the neutrino mass matrix,  we further elaborate on additional field representations within the $A_4$ symmetry. The transformations are designated as follows: $ \psi_{L,R} \sim 3$, $\nu_{1,R}\sim 1$, $\nu_{2,R}\sim 1''$, $\nu_{3,R}\sim 1'$. The flavon sector includes a singlet $\xi\sim 1'$  and two triplets, $\varphi$ and $\varphi^\prime$, with respective vevs structured as $ \langle \xi \rangle = v_\xi$, $ \langle \varphi \rangle =  (v_1,0,v_2)$ and $ \langle \varphi^\prime \rangle =  (0,v_3,0)$. Here, we do not make detailed discussion of the full flavor-symmetric scalar potential whose minimization results in these vev patterns. To impose further constraints on the Yukawa matrices, an additional $Z_4$ flavor symmetry is introduced. Under this symmetry, $\psi_{L, R}$, $\varphi$  and $\xi$ change signs; $\nu_{2,R}$ and $\varphi'$ transform as $\nu_{2,R}\to i \nu_{2,R}$ and $\varphi^\prime\to i \varphi^\prime$, respectively, while all other fields remain invariant. This symmetry configuration leads to the following forms for the neutrino mass-related Yukawa matrices:
\begin{equation}
	Y_\psi = -y_\psi \diag(1, \omega, \omega^2), \quad M=M\mathbf{I}_3, \quad Y_\nu =\begin{pmatrix}
		y_{11}  & 0 & y_{13} \\
		0 & y_{22} & 0 \\
		y_{31} & 0 & y_{33}
	\end{pmatrix},
\end{equation}	
where flavon vevs to high energy scale ratios are absorbed into  the redefined Yukawa couplings. Consequently, the effective light neutrino mass matrix is given by
\begin{equation}
	m_\nu =\frac{uvy_\psi}{2M}\begin{pmatrix}
		y_{11}  & 0 & y_{13} \\
		0 & \omega y_{22} & 0 \\
		\omega^2 y_{31} & 0 & \omega^2 y_{33}
	\end{pmatrix}.
\end{equation}
The neutrino mass matrix eigenvalues are derived by diagonalizing the Hermitian matrix $\ m_\nu  m_\nu^\dagger$ through a unitary transformation $\mathcal{U}_{L}^{(\nu)}$ acting on the left-handed neutrino fields. The eigenvalues give neutrino masses,
\begin{equation}
	\begin{aligned}\label{eq;nums}		
		m^{(\nu)}_{1,3} = & \frac{uv\left|y_\psi\right|}{2\sqrt{2}M}\left[\left|y_{11}\right|^2+\left|y_{13}\right|^2+ \left|y_{31}\right|^2+\left|y_{33}\right|^2\mp \sqrt{\triangle}\right]^{1/2}, \\
		\triangle= & \left( \left|y_{11}\right|^2+\left|y_{13}\right|^2+ \left|y_{31}\right|^2+\left|y_{33}\right|^2 \right)^2 -4\left| 
		y_{13}  y_{31} -y_{11}y_{33}\right|^2,\\
		m^{(\nu)}_2= & \frac{uv\left|y_\psi\right|}{2M}\left|y_{22}\right|. \\
	\end{aligned}	
\end{equation}
The unitary rotation matrix, $\mathcal{U}_{L}^{(\nu)}$, parameterizing the transformation is defined as
\begin{equation}
	\begin{aligned}\label{eq:nurotat}
		& \mathcal{U}_{L}^{(\nu)} =  \begin{pmatrix}
			\cos \alpha & 0 & \sin \alpha\ e^{-i \beta} \\
			0 & 1 & 0 \\  
			-\sin \alpha\ e^{i \beta} & 0 & \cos \alpha
		\end{pmatrix}.
	\end{aligned}
\end{equation}
The angular parameters $\alpha$ and $\beta$ are determined by
\begin{equation}
	\begin{aligned}\label{eq:alphabetapara}
		\tan 2\alpha = & \frac{2\left|  y_{11} y_{13}^*  + y_{31} y_{33}^*\right|}{\left|y_{13}\right|^2 + \left|y_{33}\right|^2 -\left|y_{11}\right|^2 - \left|y_{31}\right|^2},\\ 
		 \beta = & \arg \left(y_{11} y_{13}^*  + y_{31} y_{33}^* \right).
	\end{aligned}
\end{equation}

\begin{figure}
	\centering
	\includegraphics[width=0.7\linewidth]{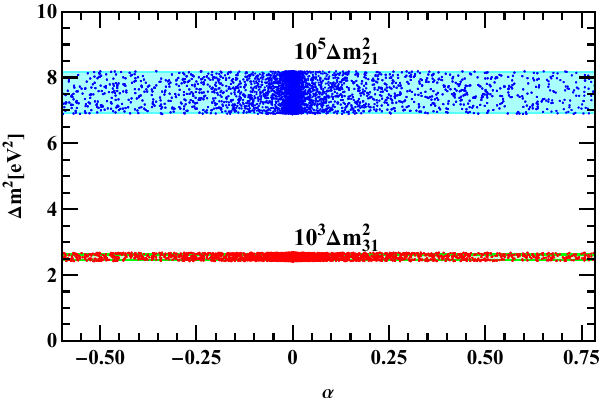}
	\caption{Scaled neutrino mass-squared differences plotted against the allowed range of $\alpha$. The solar neutrino mass-squared difference is represented by blue points and is scaled by $10^{5}$, while the atmospheric neutrino mass-squared difference, represented by red points, is scaled by $10^{3}$. The horizontal bands, cyan for solar and green for atmospheric differences, indicate the current experimentally allowed $3\sigma$ regions for neutrino normal mass ordering.}
	\label{fig:alpha_numassqr}
\end{figure}

Furthermore, beyond the dependencies on the moduli of the neutrino mass matrix elements, the neutrino masses show a correlation with the parameter $\alpha$ as encapsulated in the following relationship:
\begin{equation}
	\triangle = \frac{1}{ \cos^2 2\alpha}\left(\left|y_{13}\right|^2 + \left|y_{33}\right|^2 -\left|y_{11}\right|^2 - \left|y_{31}\right|^2 \right)^2.
\end{equation}
No similar dependency on $\beta$ is observed. As illustrated in Figure~\ref{fig:alpha_numassqr}, the neutrino mass-squared differences can be accommodated within the permissible range of $\alpha$, which is further discussed below. For this analysis, we performed a numerical scan, varying the neutrino mass matrix elements randomly within the interval $[0,0.01]$,  ensuring adherence to the mass-squared difference constraints derived from the experimentally allowed $3\sigma$ region. The results demonstrate that within the explored range of $\alpha$, the theoretical predictions for the solar and atmospheric mass-squared differences align well with the experimental constraints.

The lepton mixing matrix, which encapsulates contributions from the unitary rotations of both the charged lepton and neutrino sectors, is expressed as
\begin{equation}
	U \simeq  \left(
	\begin{array}{ccc}
		\frac{\cos \alpha - \sin \alpha \ e^{i \beta } }{\sqrt{3}} & \frac{1}{\sqrt{3}} & \frac{\cos \alpha + \sin \alpha \ e^{-i \beta } }{\sqrt{3}} \\ \\[0.01pt]
		\frac{\cos \alpha - \omega  \sin \alpha \ e^{i \beta } }{\sqrt{3}} & \frac{\omega ^2}{\sqrt{3}} & \frac{\omega  \cos \alpha + \sin \alpha \ e^{-i \beta } }{\sqrt{3}} \\  \\[0.01pt]
		\frac{\cos \alpha - \omega ^2 \sin \alpha \ e^{i \beta } }{\sqrt{3}} & \frac{\omega }{\sqrt{3}} & \frac{\omega ^2 \cos \alpha + \sin \alpha \ e^{-i \beta } }{\sqrt{3}} \\
	\end{array}
	\right).
	\label{eq:lptmixmatr}
\end{equation}
The magnitudes of the elements in this mixing matrix fit well within their $3\sigma$ experimental ranges as announced by recent studies~\cite{deSalas:2020pgw,10.5281/zenodo.4726908}. This matrix introduces two free parameters, $\alpha$ and $\beta$, while generic form of a lepton mixing matrix is parameterized by three angles and one Dirac CP phase. This reduction in parameters allows for potential predictions or correlations among the neutrino oscillation parameters. By comparing the lepton mixing matrix in eq.~\eqref{eq:lptmixmatr} with the standard parameterization~\cite{PhysRevD.110.030001}, we derived expressions that describe the functional dependencies of the neutrino mixing angles on $\alpha$ and $\beta$, as well as predictions for the Jarlskog invariant~\cite{PhysRevLett.55.1039}, which quantifies leptonic CP violation in a rephasing invariant manner: 
\begin{equation}
	\begin{aligned}\label{eq:thetasandJcp}
	&\sin^2 \theta_{12}= \frac{1}{2-\sin 2\alpha \cos \beta}, \quad 
	 \sin^2 \theta_{13} =  \frac{1}{3}\left(1+\sin 2 \alpha  \cos \beta \right) , \\
	 &\sin^2 \theta_{23}= \frac{1}{2}\left(1-\frac{\sqrt{3}\sin 2\alpha  \sin \beta}{2-\sin 2\alpha \cos \beta}\right), \quad  J_{\rm CP} = -\frac{\cos 2\alpha}{6\sqrt{3}}.
	\end{aligned}	
\end{equation}
These equations reveal correlations among the mixing angles and the CP violating phase
\begin{equation}
	\sin^2 \theta_{12}=\frac{1}{3\cos^2 \theta_{13}},\quad \tan 2\theta_{23}\cos \delta_{\rm CP} = \frac{\cos 2\theta_{13}}{\sin \theta_{13}\sqrt{2-3\sin^2 \theta_{13}}}.
\end{equation}
With known values for $\alpha$ and $\beta$, the Dirac CP violating phase is determined by 
\begin{equation}
	\cos \delta_{\rm CP}= \frac{\cos \alpha+\sin \alpha \cos \beta}{\sqrt{1+\sin2\alpha \cos \beta}}.
	\label{eq:expofcpvltphs}
\end{equation}
This indicates two special cases where leptonic CP violation is either absent or maximal. For instance, $\alpha=0$ results in $\delta_{\rm CP}=0$, regardless of $\beta$. Conversely, $\beta=0$ sets $\delta_{\rm CP}$ to $0$ or $\pi$, independent of $\alpha$.  These are the points where no leptonic CP violation appears. The condition $\cot \alpha =- \cos \beta$ implies maximal CP violation with $\delta_{\rm CP}=\pm \frac{\pi}{2}$.

\begin{figure}[htbp]
	\centering
	\subfigure[]{
		\includegraphics[scale=0.47]{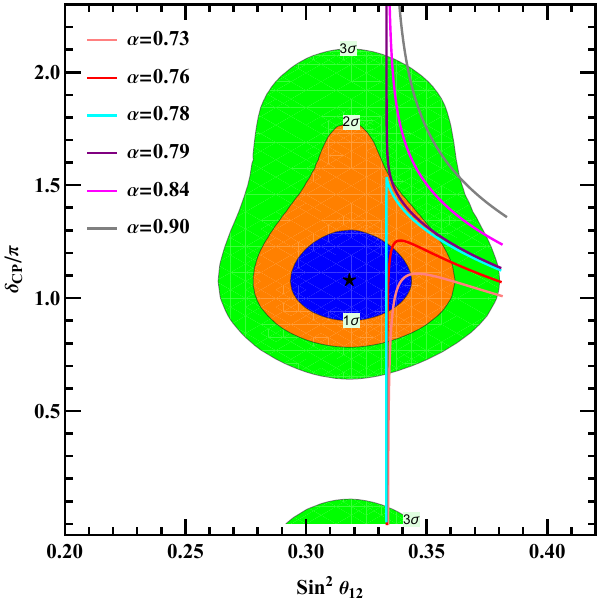} \label{subfig:S12sq_Delcp}
	}
	\subfigure[]{
		\includegraphics[scale=0.47]{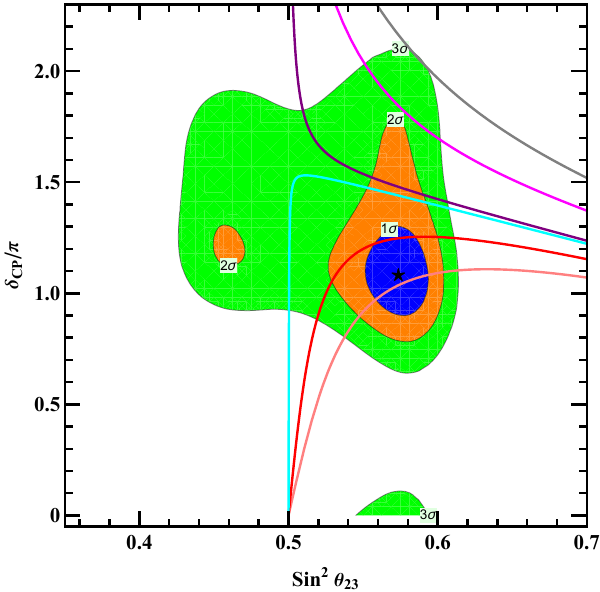} \label{subfig:S23sq_Delcp} 
	}
	\subfigure[]{
		\includegraphics[scale=0.47]{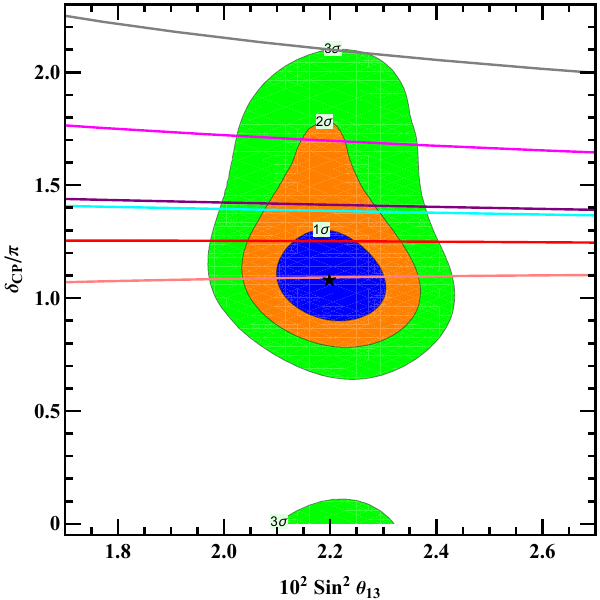}\label{subfig:S13sq_Delcp}
	}
	\caption{Plots of the CP-violating phase over $\sin^2 \theta_{12}$ (shown in (a)), $\sin^2 \theta_{23}$ (shown in (b)) and $10^2 \sin^2 \theta_{13}$ (shown in (c)), for given values of $\alpha$ indicated in the legend, which are taken the same for all plots, and range of $\beta \in [-\pi,-2.25]$. The black stars are the global best-fit points, the blue, orange and green contours are the current $1\sigma$,  $2\sigma$ and  $3\sigma$ regions, respectively, from the global fit of neutrino oscillation parameters~\cite{deSalas:2020pgw,10.5281/zenodo.4726908}.}
	\label{fig:DeltaCP_sinthetasqr}
\end{figure}

Comparative analysis of theoretical results for the Jarlskog invariant, given in eq.~\eqref{eq:thetasandJcp}, restricts $\alpha$ to a range of $-0.60 \le \alpha \le 0.92 $, which aligns with the experimental $3\sigma$ range of $-0.0348\le J_{\rm CP}\le 0.0263$~\cite{deSalas:2020pgw,10.5281/zenodo.4726908}. Additional constraints on $\alpha$ and $\beta$ stem from global fit data regarding mixing angles, CP violating phase, and their correlations. As depicted in figure~\ref{fig:DeltaCP_sinthetasqr}, the theoretical predictions are consistent with experimental results, given the same values and ranges of $\alpha$ and $\beta$. The sharp turn occur in the curves of panels (a) and (b) when $\alpha =\frac{\pi}{4}$ as it is a singular point, c.f.~eq.\eqref{eq:alphabetapara}. At this point, the denominator of~eq.\eqref{eq:expofcpvltphs} vanishes for $\beta=\pi$. The panel (c) depicts that theoretical result can cover the region which lies above the best fit point. 

\begin{figure}[htbp]
	\centering
	\subfigure[]{
		\includegraphics[scale=0.47]{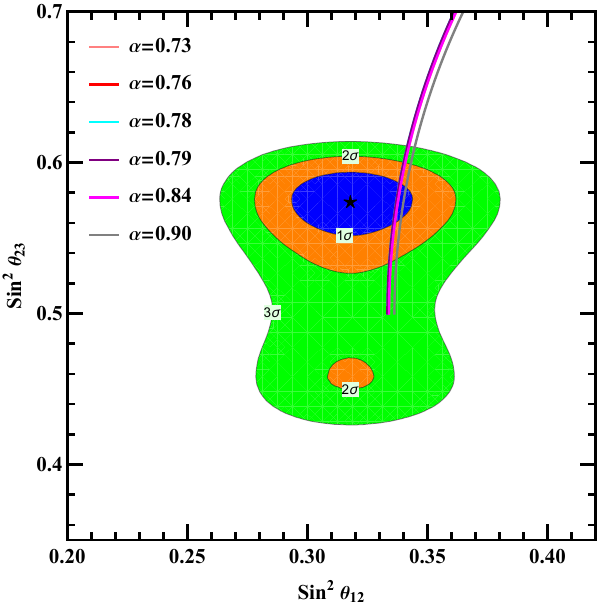} \label{subfig:S12sq_S23sq}
	}
	\subfigure[]{
		\includegraphics[scale=0.47]{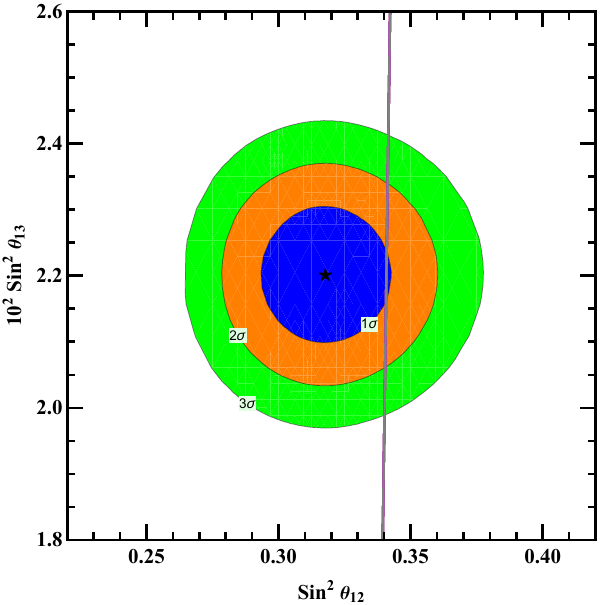} \label{subfig:S12sq_S13sq} 
	}
	\subfigure[]{
		\includegraphics[scale=0.47]{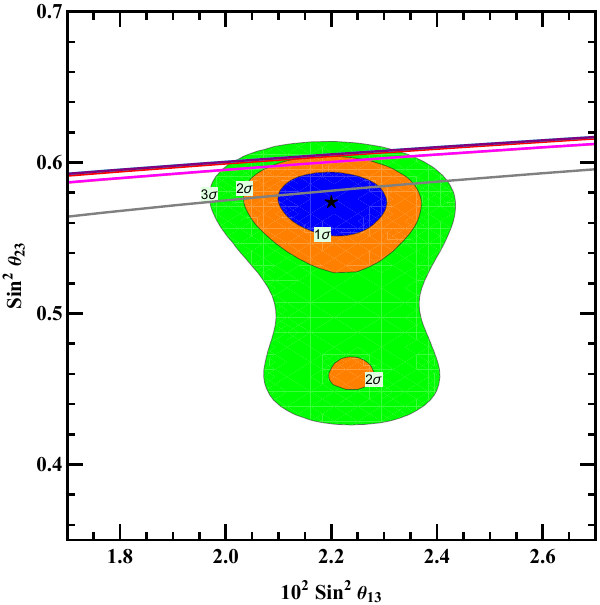}\label{subfig:S13sq_S23sq}
	}
	\caption{Correlations among the solar, atmospheric, and reactor mixing angles: (a) shows $\sin^2 \theta_{12}$ versus $\sin^2 \theta_{23}$, (b) shows $\sin^2 \theta_{12}$  versus $10^2\sin^2 \theta_{13}$, and (c) shows $10^2\sin^2 \theta_{13}$  versus $\sin^2 \theta_{23}$. Parameter values (or ranges), plotting styles, and experimentally allowed regions are consistent with those presented in figure~\ref{fig:DeltaCP_sinthetasqr}.}
	\label{fig:sinthetasqr_sinthetasqr}
\end{figure}

Figure~\ref{fig:sinthetasqr_sinthetasqr} further demonstrates how the chosen parameter space can successfully accommodate additional correlations among the sine squares of the mixing angles, providing a good match with experimental constraints across the $1-3\sigma$ regions. Panel (a) of the figure illustrates the correlation between $\sin^2 \theta_{12}$ and $\sin^2 \theta_{23}$. Notably, the lower boundaries of the curves are constrained by the upper limit of the parameter range for $\beta$, which is established based on other correlated observations within the model. The panel (b) depicts the relationship between $\sin^2 \theta_{12}$ and $10^2 \sin^2 \theta_{13}$. Interestingly, despite varying $\alpha$ and $\beta$ within their respective ranges, all curves overlap due to both $\sin^2 \theta_{12}$ and  $\sin^2 \theta_{13}$ being dependent on the single parametric combination of $\sin2\alpha \cos \beta$, as formulated in eq.~\eqref{eq:thetasandJcp}. Although taking different values for $\alpha$ and changing $\beta$ in its range, the combination remains the same. This dependency causes all theoretical curves to merge into a single line. Notably, the plots in panel (b) of figure~\ref{fig:DeltaCP_sinthetasqr} and panel (c) of figure~\ref{fig:sinthetasqr_sinthetasqr} reveal that $\theta_{23}$ predominantly resides within the higher octant. This observation is particularly significant given the ongoing experimental ambiguity in determining the precise octant of the atmospheric mixing angle.

\subsection{Dark matter relic density}
\label{subsec:DMphnmlgy}

In addition to elucidating the mechanism for neutrino mass and mixing, this model introduces a DM candidate crucial for neutrino mass generation. The neutral component of the vector-like lepton, interacting solely through weak interactions, has its lightest right-handed component potentially long-lived. Given that the scalar $\phi$ is heavier than the lightest right-handed component of the lepton, $m_\phi > m_0$, (where $m_0$ is the mass of the lightest $\psi^0_{i,R}$), kinematic constraints prevent two-body decay processes, although the neutral component does couple to the neutrino and scalar $\phi$. An additional two-body decay channel $\psi^0_R \to \nu_L h$ may be facilitated through the Yukawa interaction $Y_\psi$ and scalar mixing. The corresponding decay rate is given by
\begin{equation}
    \Gamma_2 = \frac{\left|Y_\psi\right|^2 \lambda^2 u^2 v^2 }{64\pi m^3_0}\left( \frac{m^2_0-m^2_h}{m^2_\phi-m^2_h}\right)^2,
    \label{eq:psirdhnv}
\end{equation}
where the neutrino mass is considered negligible and the scalar mixing angle is approximated by $\sin \theta \approx \frac{\lambda u v }{2(m^2_\phi - m^2_h)}$. As discussed previously, both the Yukawa coupling $Y_\psi$, which characterizes the mixing between the SM and newly introduced leptons, and $\lambda$, representing scalar mixing, are very small. Moreover, the decay rate is further suppressed by the heavy masses in the denominator, suggesting that the decay proceeds very slowly and $\psi^0_R$ may be long-lived.

\begin{figure}[htbp]
\centering
\subfigure[ ]{\includegraphics[width=7cm]{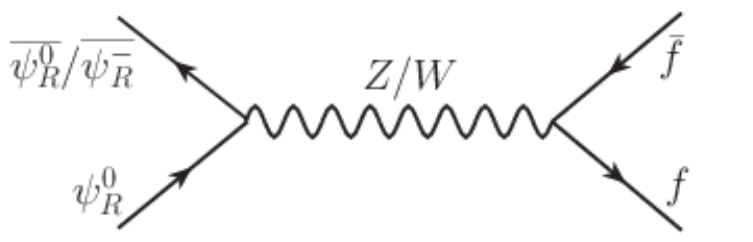}}
\subfigure[ ]{\includegraphics[width=7cm]{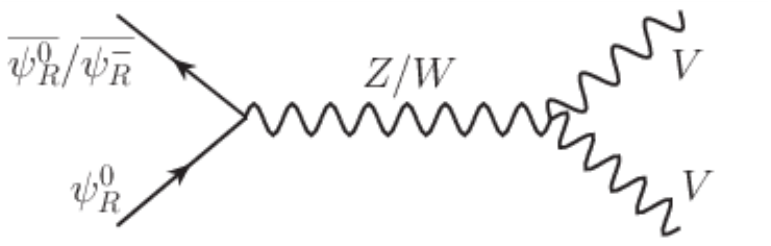}}
\subfigure[ ]{\includegraphics[width=7cm]{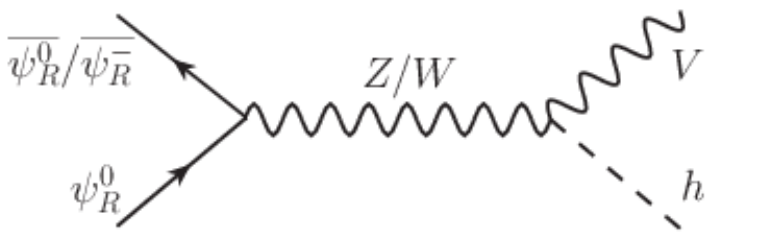}}
\subfigure[ ]{\includegraphics[width=7cm]{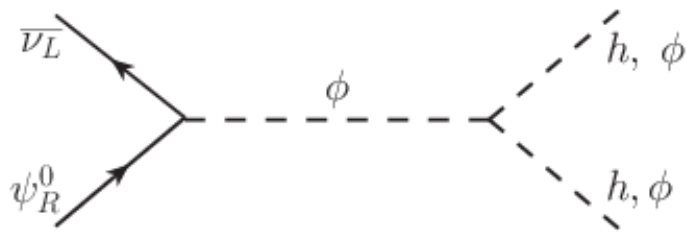}}
\caption{Tree-level processes contributing to the annihilation of the WIMP DM, where $f$ denotes the SM fermions, and $V$ represents the vector bosons which include weak gauge bosons $W$, $Z$, or photon.}
\label{fig:feynmannwimpanhltn}
\end{figure}

Expanding upon this, we explore the three-body decay process $\psi^0_{R}\to \nu h h$, with $\phi$ acting as a mediator. The decay rate for this process is described by the following equation:
\begin{equation}
\Gamma_3 = \frac{\left|Y_\psi\right|^2 \lambda^2}{(16\pi)^3} \frac{\cos^4 \theta u^2(m_0-2m_h)}{m_0^2}\left[ \frac{m_0^2(m_0-2m_h)}{m_0^2 m_h -m_0 m_\phi^2 +m_h m_\phi^2}+\ln \frac{m_0 m_\phi^2-m_h m_\phi^2 -m_h m_0^2}{(m_0- m_h) ( m_\phi^2-m_0^2) } \right],
\end{equation} 
where neutrino mass is neglected. The small couplings $Y_\psi$ and $\lambda$, combined with the heavy mass of the decaying particle, result in an extremely low decay rate, extending its potential lifetime beyond the age of the universe. As this particle participates only in the weak interaction, this characteristic classifies it as a Weakly Interacting Massive Particle (WIMP). The relevant annihilation processes of this WIMP are depicted in figure~\ref{fig:feynmannwimpanhltn}.

The evolution of the number density $n$ of DM particle in the expanding universe is described by the Boltzmann equation,
\begin{equation}
    \frac{dn}{dt} + 3H n = - \langle \sigma v \rangle \left( n^2 - n_{\text{eq}}^2 \right),
\end{equation}
where $ H$  is the Hubble expansion rate, $n_\chi^{\text{eq}}$ is the equilibrium number density, and  $ \langle \sigma v \rangle$  is the thermal average of the annihilation cross section times the relative velocity, detailed calculations of which are found in Appendix~\ref{app:DMrelicden}. It is a standard approach to introduce DM yield $Y$ in order to simplify the equation and make it more suitable for cosmological calculations. The yield is defined by the ratio, $Y=n/s$, which is the number density normalized by the entropy density $s(T)= \frac{2\pi^2}{45} g_{*s} T^3$.
Due to the conservation of entropy per comoving volume, the Boltzmann equation simplifies to
\begin{equation}
    \frac{dY}{dx} = - \frac{\langle \sigma v \rangle s}{H x} \left( Y^2 - Y_{\text{eq}}^2 \right),
    \label{eq:bltzmaneq}
\end{equation}
with the variable change from time $t$ to the dimensionless quantity $x=m_0/T$. In the non-relativistic regime $x\gg 1$, $Y_{\text{eq}}$ is suppressed by the exponential of $x$, thus can be neglected when considering the late time evvolution of the cold DM. As the universe expands, the temperature decreases, causing the DM interaction rate to drop below the expansion rate, leading to decoupling from the thermal plasma and freeze-out of the comoving density. Freeze-out occurs when the annihilation rate equals the expansion rate, which is quantified by 
 \begin{equation}
     n_{\text{eq}} \langle \sigma v \rangle = \frac{2m^2_0\pi}{3x^2M_{\text{Pl}}}\sqrt{\frac{\pi}{5}g_*}\, ,
 \end{equation}
defining the freeze-out temperature $T_f$ or $x_f = m_0/T_f$ with typical values around $x_f \sim \mathcal{O}(10)$. Integrating both sides of the simplified Boltzmann equation from the freeze-out time $x=x_f$ to late times $x=\infty$, we obtain
\begin{equation}
    Y^{-1}_\infty \approx \int_{x_f}^{\infty} \frac{\langle \sigma v \rangle s}{H x} dx,
\end{equation}
where the $1/Y_f$ term is neglected due to its insignificance compared to late-time values. Using expressions for $s$ and $H$,
\begin{equation}
    \begin{aligned}
        s(x)& = \frac{2\pi^2 g_{*s} m^3_0}{45x^3},\\
        H(x)&= \frac{2\pi m_0^2}{3M_{\text{Pl}}x^2}\sqrt{\frac{\pi g_*}{5}},
    \end{aligned}
\end{equation}
and performing the integration leads to
\begin{equation}
\begin{aligned}
        Y_\infty \approx  \frac{9 }{m_0 M_{\text{Pl}} g_{*s}} \sqrt{\frac{5g_*}{\pi}}\bigg[ & \frac{g^4}{m^2_0} 2.72\times 10^{4} +\left(\frac{1}{u^2} 1.82\times 10^{-250} \right. \\
    & \left. \left. +\frac{\lambda^2}{m^4_0}\left(3.02 u^2 +0.13 v^2    \right)\times 10^{-59} \right) \left|Y_\psi\right|^2 \cos^2\theta\right]^{-1},
\end{aligned}
\end{equation}
where $m_\phi=5m_0$ and $x_f =12$ are used. Given that terms beyond the first in the square bracket are significantly suppressed, they are neglected in subsequent discussions. The relic abundance of WIMP DM is thus described by
\begin{equation}
	\Omega h^2 = \frac{Y_\infty s_0 m_0}{\rho_c/h^2} \approx  \frac{4.18 \times 10^{-4} \sqrt{g_*} m^2_0}{g_{*s}g^4 M_{\text{Pl}} \rho_c/h^2  }.
	\label{eq:DMrelicabundnce}
\end{equation}
Upon substituting the numerical values into our calculations, we obtain a clearer picture of the DM scenario. With the number of effective entropy degrees of freedom $g_* \approx g_{*s}\sim 100$, the weak gauge coupling $g=0.65$, the current entropy density $s_0 =2891~\text{cm}^{-3}$, the reduced critical energy density $\rho_c/h^2 = 1.0537 \times 10^{-5} ~ \text{GeV cm}^{-3}$, and the Planck mass $M_{\text{Pl}}=1.22\times 10^{19} ~\text{GeV}$,  the relic density formula simplifies to $\Omega h^2 = 0.05265 (m_0/\text{TeV})^2$. For the DM mass $m_0=1.51 ~\text{TeV}$, the calculated relic density is $\Omega h^2 =0.120$, which aligns with the DM relic density $\Omega h^2 = 0.120\pm 0.001$ observed by the Planck collaboration~\cite{Planck:2018vyg}. This congruence validates our theoretical framework and the assumptions made about the properties of the DM particle under consideration.

\subsection{Lepton flavor violation and exotic lepton searches}
\label{subsec:LFV&HNL}

In the SM with massless neutrinos, individual lepton numbers are conserved as accidental symmetries. The observation of neutrino oscillations, however, confirms that neutrinos are massive and that these symmetries are not exact. Furthermore, charged-lepton flavor violation (LFV) is a generic prediction of SM extensions. LFV processes are typically probed through rare processes, such as $\ell_\alpha \to \ell_\beta \gamma$, $\ell_\alpha \to \ell_\beta \bar{\ell}_\mu \ell_\nu$, and conversion of $\mu^- \to  e^-$ in nuclei. The first proceed with loop decay, the second is more sensitive to four-lepton flavor-changing contact interactions, and the last one is due to a new interaction between the leptons and quarks. Concerning relevance to our work, here we perform a detailed analysis on two representative channels $\mu\to e \gamma$ and $\mu \to eee$. 

\begin{figure}[!ht]
	\centering
	\includegraphics[width=0.5\linewidth]{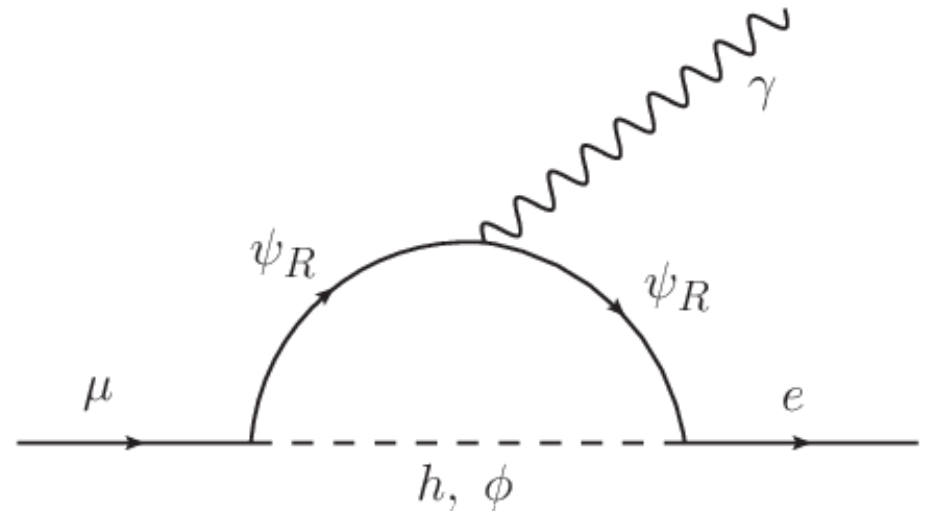}
	\caption{New physics contribution to $\mu\to e \gamma$.}
	\label{fig:mu2egfyn}
\end{figure}

In addition to the loop contribution mediated by the $W$ boson and neutrino mixing, Yukawa couplings of the Lagrangian in eq.~\eqref{eq:yukawalag} generate additional contributions to the $\mu \to e \gamma$ process through the loop diagram shown in Figure~\ref{fig:mu2egfyn}. 
The branching ratio of this process is 
\begin{equation}
	\begin{aligned}
		\text{BR}(\mu \to e\gamma) = \frac{3\alpha}{16\pi G_F^2 m_\mu^2} \sum_{i} & \left|  \frac{ G_F m_\mu}{\sqrt{2}}  U_{\mu i} U_{e i}^* \, \frac{m_{\nu_i}^2}{m_W^2} - Y_{\psi, \mu i} Y_{\psi, e i}^*  \frac{m_{\psi_i}}{m_\phi^2} \cos^2\theta f_1\left(\frac{m_{\psi_i}^2}{m_\phi^2}\right)\right. \\
		& \quad \left.  + Y_{\psi, \mu i} Y_{\psi, e i}^* \frac{m_h}{m_{\psi_i}^2} \sin^2\theta f_2\left(\frac{m_h^2}{m_{\psi_i}^2}\right)\right|^2,
	\end{aligned}
\end{equation}
where $\alpha$ is the fine structure constant, $G_F$ is the Fermi constant, and $f_{1, 2}(x)$ are the loop functions:
\begin{equation}
	f_1(x) = \frac{1-x+ \ln x }{\left(1-x\right)^2 }, \quad 	f_2(x) = f_1(x)-\frac{\ln x}{1-x}.
\end{equation}

\begin{figure}[!ht]
	\centering
	\includegraphics[width=0.8\linewidth]{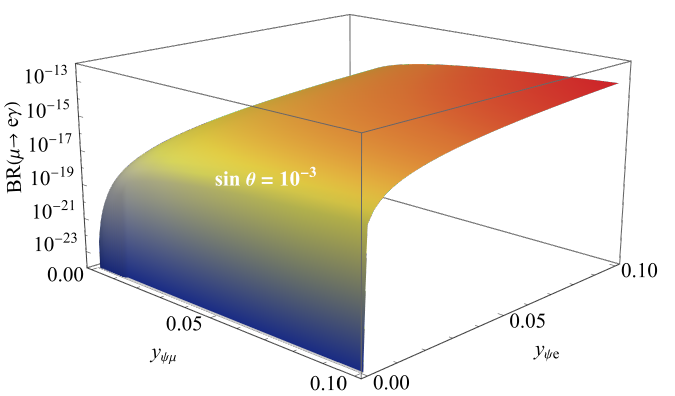}
	\caption{Branching ratio of the $\mu \to e\gamma$ process as a function of Yukawa couplings.}
	\label{fig:clfv}
\end{figure}

To extract the main features of our theoretical result, we performed a simplified numerical analysis. Given the parameter region $m_{\psi_i} =1.51$ TeV and $m_\phi = 5 m_{\psi_i}$ required to explain the DM relic density, we consider universal flavor couplings of all three generations of vector-like leptons to muon and electron, $Y_{\psi, \mu i}=y_{\psi\mu}$ and $Y_{\psi, e i}=y_{\psi e}$. With a fixed scalar mixing $\sin \theta = 10^{-3}$, figure~\ref{fig:clfv} shows the resulting branching ratio for $\mu^+ \to e^+ \gamma$ as a function of these couplings. Although the branching ratio grows with increasing coupling, it remains well below the current most stringent experimental limit from MEG II, $\mathrm{Br\left(\mu^+ \to e^+ \gamma  \right)}< 3.1 \times 10^{-13}$~\cite{MEGII:2023ltw}. 
We also examine the impact of scalar mixing and observe that the branching ratio exhibits only weak dependence on the mixing angle, showing a slight decrease as the angle increases. For parameter values of $\sin \theta = y_{\psi e}=y_{\psi \mu}=0.1$, we obtain $\mathrm{Br\left(\mu^+ \to e^+ \gamma  \right)=5.4\times 10^{-14}}$, which is one order of magnitude below the current MEG II upper limit. 

The decay $\mu \to eee$ receives contributions from dipole, self-energy, and box diagrams.  Given that the dipole operator provides the dominant contribution, the branching ratios for $\mu \to eee$ and $\mu \to e\gamma$ are correlated~\cite{Kuno:1999jp}, 
\begin{equation}
   \text{BR}(\mu \to eee) \simeq\frac{\alpha}{3\pi}\left(\ln\frac{m_\mu^2}{m_e^2}-\frac{11}{4}\right) \text{BR}(\mu \to e\gamma) .
\end{equation}
Using this relation, we obtain an upper limit of $\text{BR}(\mu \to eee) \le 3.4\times 10^{-16}$, which lies four orders of magnitude below the current experimental bound of $\text{BR}(\mu \to eee) < 1.0 \times 10^{-12}$~\cite{SINDRUM:1987nra}. This rate may be accessible to future experiments such as Mu3e.

As discussed in the previous sections, the neutral components of the vector-like leptons are heavy, weakly interacting particles, with the lightest right-handed state potentially serving as a DM candidate. Compared to the right-handed components of the vector-like lepton, the decay rate for $\psi^0_L \to \nu_R h$ while formally similar to that in eq.~\eqref{eq:psirdhnv}, does not get suppressed from scalar mixing. scalar mixing. We now turn our attention to their collider signatures. Both left- and right-handed neutral components mix with neutrinos following the breaking of extended and electroweak symmetries, but their masses are all above the TeV scale and remain consistent with LHC constraints~\cite{CMS:2024hik,ATLAS:2025zmk}. Moreover, the charged leptons mix with the vector-like states, contributing to LFV processes. In addition, considering both gauge interactions and Higgs couplings, several distinctive collider signatures may emerge, such as Drell-Yan pair production $pp \to Z^* \to \overline{\psi^0} \psi^0$, associated production with SM gauge bosons $pp \to W^\pm \to \psi^\pm \psi^0$, and Higgs-mediated processes $pp \to h \to \overline{\psi^0} \nu$. Current LHC searches set a lower mass bound of approximately 1 TeV for these vector-like leptons~\cite{CMS:2022nty,CMS:2024bni}. Future collider experiments will further probe the mass ranges of these exotic leptons in our model.

The scalar sector of this model includes a heavy scalar $\phi$, composed mainly of the singlet scalar $\sigma$. As a gauge singlet, $\phi$ has no direct couplings to SM quarks or gauge bosons, interacting mainly through Higgs mixing with suppressed strength. Its dominant couplings involve neutrinos, charged leptons, and vector-like leptons\footnote{Here, all fields are written in mass basis.},
\begin{equation}
	\mathcal{L}_{\rm Y} \supset  -\cos \theta  Y_{\psi,ij}  \left[\left(\mathcal{U}_{L}^{(\nu)*}\right)_{ik} \overline{\nu_{k L}}  \psi^0_{j,R}+\left(U^{(e)*}_L\right)_{ik}\overline{e_{k L}}  \psi^-_{j,R}\right]  \phi,
\end{equation}
which is very weak. These suppressed couplings make $\phi$ particularly challenging to produce in high-energy colliders. Furthermore, if it has a sizable coupling to Higgs, this leads to decays of $\phi$ into Higgs pairs, making it difficult to detect in high-energy collider environments due to its fleeting presence.

\section{Summary and discussion}
\label{sec:conclusions}

We have presented a minimal realization of the Dirac seesaw mechanism through an extension of the SM symmetry by discrete $Z_2\times Z_3$ groups, accompanied by expanded scalar and leptonic sectors. The $Z_2$ symmetry distinguishes SM fields from newly introduced fields, while the unbroken $Z_3$ symmetry enforces an effective lepton number conservation, ensuring the Dirac nature of neutrinos by forbidding Majorana mass terms. This framework successfully accommodates the observed neutrino normal mass ordering, mixing angles, and leptonic CP-violating phase, and provides a viable WIMP DM candidate.

The extended particle content and flavor-violating couplings naturally give rise to testable phenomenological signatures. Our analysis of LFV processes reveals consistency with current experimental constraints, the rates for $\mu \to e\gamma$ and $\mu \to eee$ within the sensitivity of next generation experiments. The TeV-scale HNL as well as vector-like leptons, though consistent with the LHC limits, will be further probed through Drell-Yan production, associated production, and Higgs-mediated processes at future colliders.

The results of the model align with existing experimental observations while remaining falsifiable through upcoming experiments. A detection of $0\nu\beta\beta$ decay would confirm the Majorana nature of neutrinos, thereby ruling out our Dirac neutrino framework. Experiments such as KamLAND-Zen~\cite{KamLAND-Zen:2024eml}, LEGEND-1000~\cite{LEGEND:2021bnm}, nEXO~\cite{nEXO:2021ujk}, and CDEX-300$\nu$~\cite{CDEX:2022bdk} are poised to explore the parameter space for both inverted and normal mass orderings. The model preference for neutrino normal mass ordering will be tested by JUNO~\cite{JUNO:2024jaw} and DUNE~\cite{DUNE:2015lol}.  
 The indication of the upper octant for $\theta_{23}$ and the determination of the leptonic CP-violating phase $\delta_{CP}$ will be tested by long-baseline experiments, including DUNE~\cite{DUNE:2015lol} and Hyper-Kamiokande~\cite{SajjadAthar:2021prg}.  Future searches for $\mu \to e \gamma$ (MEG II) and $\mu \to e e e$ (Mu3e) will further constrain the Yukawa couplings and the mass scale of the model.  

The introduction of additional scalars and fermions opens possibilities for new phenomena. In particular, the heavy-mass scale and complex Yukawa structures could facilitate Dirac leptogenesis, offering a pathway to explain the matter-antimatter asymmetry of the universe. Although our construction uses discrete symmetries for simplicity, a natural extension would involve embedding $Z_2\times Z_3$ into larger continuous or discrete symmetry groups. Such generalizations may yield richer phenomenology and provide a more fundamental realization of the Dirac seesaw mechanism. These directions, though beyond the present scope, are slated for investigation in a future study.

\section*{Acknowledgements}
The work of Y.R. is supported by the Natural Science Foundation of Xinjiang Uygur Autonomous Region of China under Grant No. 2022D01C52. The work of M.A. is supported by the National Natural Science Foundation of the People’s Republic of China (No. 12303002) and Tianchi talent project of Xinjiang Uygur Autonomous Region of China.


\appendix

\section{Diagonalization of lepton mass matrices}
\label{app:Dialepmas}
In this appendix, we detail the diagonalization procedure for the lepton mass matrix, from which the physical masses and mixing parameters are derived. First of all, the mass matrix in the charged lepton sector 
\begin{equation}\label{eq:MEmatr}
	M_E= \begin{pmatrix}
		m_E &  m_\psi \\
		0     &  M
	\end{pmatrix},
\end{equation}
can be block-diagonalized as
\begin{equation}\label{eq:bdgnlz}
	\tilde{U}_L^\dagger M_E \tilde{V}_R = \left(	\begin{array}{c:c}
	\qquad \hspace{12pt} & \qquad \\[-5pt]
	 m_E  & \mathbf{O}_{3} \\[3pt] \hdashline
	\qquad \hspace{12pt} & \qquad \\[-5pt]
	\mathbf{O}_{3}  & M +\frac{1}{2}x^\dagger m_\psi  \\[3pt] 
	\end{array} \right),
\end{equation}
in leading orders of small $3 \times 3$ dimensionless perturbation matrices $ x \equiv \left(m_\psi M^{-1}\right)^*$ and $ z \equiv m_E \left(M^{-1} x^\dagger \right)^T$, assuming the block matrix $M$ is non-singular. Two rotation matrices in this diagonalization are 
\begin{equation}
	\tilde{U}_L=\begin{pmatrix}
		\mathbf{I}_{3} - \frac{1}{2}x  \, x^T & x\\[5pt]
		- x^T & \mathbf{I}_{3} - \frac{1}{2}x^T\, x 
	\end{pmatrix}, \quad 
		\tilde{V}_R=\begin{pmatrix}
		\mathbf{I}_{3} - \frac{1}{2} z \, z^T & z  \\[5pt]
		-z^T & \mathbf{I}_{3} - \frac{1}{2}z^T \, z 
	\end{pmatrix},
\end{equation}
where $\mathbf{I}_{3}$ and $\mathbf{O}_{3}$ are three dimensional identity matrix and zero matrix, respectively. Furthermore, each blocks in eq.~\eqref{eq:bdgnlz} are further diagonalized by bi-unitary rotations, 
\begin{equation}
	\begin{aligned}
		& U_{L}^{(e)\dagger} m_E V^{(e)}_{R} = \diag(m_{1}, m_{2},m_{3}), \\
		& U_{L}^{(\psi)\dagger} \left(M +\frac{1}{2}x^\dagger\,  m_\psi\right) V^{(\psi)}_{R} = \diag(M_{1}, M_{2}, M_{3}).
	\end{aligned}
\end{equation}
Here, the notations $m_i$ and $M_i$, respectively, denote the charged lepton and the vector-like lepton masses. The complete diagonalization of the mass matrix in eq.~\eqref{eq:MEmatr} is achieved by
\begin{equation}\label{eq:flldiag}
	U_L^\dagger M_E V_R = \diag(m_i, M_i),
\end{equation} 
where $U_L = \tilde{U}_L \left(U_{L}^{(e)}\oplus U_{L}^{(\psi)}\right)$ and $V_R =\tilde{V}_R \left(V_{R}^{(e)}\oplus V_{R}^{(\psi)}\right)$.

As for the neutral lepton sector, their mass matrix is
\begin{equation}
	M_\nu= \begin{pmatrix}
		0 &  m_\psi \\
		m_D    &  M
	\end{pmatrix}.
\end{equation}
The diagonalization procedures similar to those outlined in eqs.~\eqref{eq:bdgnlz}-\eqref{eq:flldiag} are applied here, and summarized as follows. The initial block diagonalization
\begin{equation}\label{eq:numasblk}
	\tilde{\mathcal{U}}^\dagger_L M_\nu \tilde{\mathcal{V}}_R = \left(
	\begin{array}{c:c}
		\qquad & \qquad \\[-5pt]
		-m_\psi M^{-1}m_D & \mathbf{O}_{3} \\[3pt] \hdashline
		\qquad & \qquad \\[-5pt]
		\mathbf{O}_{3} & M+\frac{1}{2}m_D \left(M^{-1}m_D\right)^T  \\[3pt] 
	\end{array} \right),
\end{equation}
is facilitated by rotation matrices $\tilde{\mathcal{U}}_L$ and $\tilde{\mathcal{V}}_R$ with small correction terms $x=\left(m_\psi M^{-1}\right)^*$ and $\eta =\left(M^{-1} m_D \right)^T$, detailed as
\begin{equation}
	\tilde{\mathcal{U}}_L  =\begin{pmatrix}
		\mathbf{I}_{3} - \frac{1}{2} x\, x^T & x \\[5pt]
		- x^T & \mathbf{I}_{3} - \frac{1}{2} x^T x
	\end{pmatrix}, \quad 
	\tilde{\mathcal{V}}_R=\begin{pmatrix}
		\mathbf{I}_{3} - \frac{1}{2} \eta \, \eta^T & \eta  \\[5pt]
		-\eta^T & \mathbf{I}_{3} - \frac{1}{2}\eta^T \eta 
	\end{pmatrix}.
\end{equation}
The top-left block in eq.~\eqref{eq:numasblk} contains the light neutrino mass matrix
\begin{equation}\label{eq:lghtnumas}
	m_\nu=-m_\psi M^{-1}m_D.
\end{equation}
Each block in eq.~\eqref{eq:numasblk} is subsequently diagonalized by $3\times 3$ unitary matrices $\mathcal{U}_{L}^{(\nu)}$, $\mathcal{V}^{(\nu)}_{R}$, $\mathcal{U}_{L}^{(\psi)}$, $\mathcal{V}^{(\psi)}_{R}$ that rotate the left- and right-handed neutral leptons, yielding
\begin{align}\label{eq:numsdiag}
		&\mathcal{U}_{L}^{(\nu)\dagger} m_\nu \mathcal{V}^{(\nu)}_{R} = \diag\left(m^{(\nu)}_{1}, m^{(\nu)}_{2},m^{(\nu)}_{3}\right),\\
		& \mathcal{U}_{L}^{(\psi)\dagger} \left[M+\frac{1}{2}m_D \left(M^{-1}m_D\right)^T\right] \mathcal{V}^{(\psi)}_{R} = \diag\left(M^{(0)}_{1}, M^{(0)}_{2}, M^{(0)}_{3}\right).
\end{align}
Masses of the light neutrinos and the neutral components of the vector-like leptons are achieved by full diagonalization of the mass matrix
\begin{equation}
\begin{aligned}
	&\mathcal{U}_L^\dagger M_\nu \mathcal{V}_R = \diag \left(m^{(\nu)}_i, M^{(0)}_i\right), \\
		&\mathcal{U}_L  = \tilde{\mathcal{U}}_L \left(\mathcal{U}_{L}^{(\nu)}\oplus \mathcal{U}_{L}^{(\psi)}\right), \quad \mathcal{V}_R =\tilde{\mathcal{V}}_R\left(\mathcal{V}_{R}^{(\nu)}\oplus \mathcal{V}_{R}^{(\psi)}\right).
\end{aligned}
\end{equation}

Finally, the neutrino mixing matrix consists of the rotations of left-handed the charged and neutral lepton sectors,
\begin{equation}
    U \simeq U_{L}^{(e)\dagger} \mathcal{U}_{L}^{(\nu)}.
\end{equation}

\section{Computation of thermally averaged cross section}
\label{app:DMrelicden}

This appendix details the computation of the thermally averaged annihilation cross section. We begin by calculating the cross sections for each type of process depicted in Figure~\ref{fig:feynmannwimpanhltn}.

The cross section for the process shown in $(a)$ is given by
\begin{equation}
    \begin{aligned}
    \sigma(a)= &\frac{g^4}{96\pi}\bigg\{ \frac{1}{2\left( s-m^2_W\right)^2} \sqrt{\frac{\lambda(s,m^2_{\bar{f}}, m^2_f)}{\lambda(s,m^2_-, m^2_0)}}\bigg[2s -\left(m^2_-+m^2_0+m^2_{\bar{f}}+m^2_f\right)
     -\frac{1}{s}\bigg( m^4_-+m^4_0 + \\
    &\left. \left( m^2_{\bar{f}}-m^2_f\right)^2-2m^2_0\left( m^2_{\bar{f}}+m^2_f\right)-2m^2_-\left( m^2_0+m^2_{\bar{f}}+m^2_f\right) \right)\\
    &-\frac{1}{s^2}\bigg(\left(m^2_-+m^2_0 \right)\left(m^4_{\bar{f}}+m^4_f\right)+\left(m^2_--m^2_0 \right)^2\left(m^2_{\bar{f}}+m^2_f\right) -2m^2_{\bar{f}}m^2_f\left( m^2_-+m^2_0\right) \bigg)
        \end{aligned}
\end{equation}
\begin{equation*}
    \begin{aligned}
     & +\frac{2}{s^3}\left(m^2_--m^2_0 \right)^2\left(m^2_{\bar{f}}-m^2_f\right)^2 \bigg] +\frac{1}{\cos^4\theta_W\left( s-m^2_Z\right)^2s}\sqrt{\frac{s-4m^2_f}{s-4m^2_0} }\bigg[ \\
     & g^2_+ \biggl( s^2-s\left(m^2_0+m^2_f \right) +4m^2_0m^2_f\biggr)
     \left. -g^2_-m^2_f \left( s-2m^2_0\right) \bigg] \right\},
    \end{aligned}
\end{equation*}
where $\sqrt{s}$ is the center of mass energy, $m_-$ is the mass of $\psi^-$, the K\"all\'en function and couplings-squared are defined as 
\begin{equation}
    \begin{aligned}
        \lambda(a, b, c) = & a^2+b^2+c^2-2\left( ab+ac+bc\right),\\
        g^2_{\pm} = &   \left( g^f_V \right)^2\pm \left(g^f_A\right)^2 ,
    \end{aligned}
\end{equation}
with $Z$ boson vector and axial vector couplings $g^f_V$, $g^f_A$ to the SM fermions. 

The cross section for the process in $(b)$ is
\begin{equation}
    \begin{aligned}
        \sigma(b)= & \frac{g^2 e^2}{192\pi\sqrt{\lambda(s,m^2_-,m^2_0)}(s-m^2_W)}\bigg\{4s -5\left(m^2_-+m^2_0\right) -11m^2_W 
         +\frac{1}{s}\left[\left(m^2_--m^2_0\right)^2 \right.\\ 
         &+7\left(m^2_-+m^2_0\right)m^2_W + 4m^4_W\Big] 
         +\frac{4m^2_W}{s^2} \left[ (m^2_--m^2_0)^2+(m^2_-+m^2_0)m^2_W \right] \\
         & -\frac{8 m^4_W}{s^3}(m^2_--m^2_0)^2 \bigg\} \\
        & + \frac{g^4\cos^2 \theta_W}{192\pi s^3\left(s-m^2_W\right)^2}\sqrt{\frac{\lambda(s,m^2_W,m^2_Z)}{\lambda(s,m^2_-,m^2_0)}} \bigg\{ m^4_-\left[  s^2 +4s\left( m^2_W+m^2_Z \right) -8\left( m^2_Z-m^2_W \right)^2\right] \\
        & -m^2_-\left[ s \left[ 5s^2 -7s\left( m^2_W+m^2_Z\right)-4\left( m^2_Z-m^2_W \right)^2 \right] \right.
         +2m^2_0 \left[ s^2 +4s\left( m^2_W+m^2_Z\right) \right. \\
         & \left. \left. -8\left( m^2_Z-m^2_W \right)^2 \right] \right] + \left( s-m^2_0\right)\left[s\left[4s^2 -11s\left( m^2_W+m^2_Z\right)+4\left( m^2_Z-m^2_W \right)^2\right]  \right. \\
        & -m^2_0\left.  \left[ s^2 +4s\left( m^2_W+m^2_Z\right)-8\left( m^2_Z-m^2_W \right)^2 \right] \right]\bigg\} 
            \end{aligned}
\end{equation}
\begin{equation*}
        \begin{aligned}
        &  + \frac{g^4}{384\pi\left( s-m^2_Z\right)^2}\sqrt{\frac{\lambda(s,m^2_W,m^2_W)}{\lambda(s,m^2_-,m^2_0)}} \bigg\{4s-5\left( m^2_-+m^2_0 \right) -22 m^2_W \\
         & + \frac{8m^2_W}{s^2}\left( m^2_- - m^2_0 \right)^2+\frac{1}{s}\left[ \left( m^2_- - m^2_0 \right)^2 +14m^2_W\left( m^2_- + m^2_0 \right) \right] \bigg\}.
    \end{aligned}
\end{equation*}

The cross sections for processes $(c)$ and $(d)$ are
\begin{equation}
    \begin{aligned}
        \sigma(c)  = & \frac{g^2 m^2_W}{64\pi s (s-m^2_W)^2}\sqrt{\frac{\lambda(s,m^2_W,m^2_h)}{\lambda(s,m^2_-,m^2_0)}} \left(s-m^2_--m^2_0\right)\\
        & +\frac{g^2 m^2_Z}{128\pi \cos^4\theta_W s (s-m^2_Z)^2}\sqrt{\frac{\lambda(s,m^2_Z,m^2_h)}{\lambda(s,m^2_-,m^2_0)}} \left(s-m^2_--m^2_0 \right),
    \end{aligned}
\end{equation}
\begin{equation}
\begin{aligned}
    \sigma(d)=\frac{\left| Y_\psi \right|^2\cos^2\theta}{2^9\pi \left( s-m^2_\phi\right)^2}\left( 9\lambda^2_\sigma u^2 \sqrt{1-\frac{4m^2_\phi}{s}}
     +\frac{\lambda^2 v^2}{s} \sqrt{\lambda(s,m^2_\phi, m^2_h)}+\lambda^2 u^2 \sqrt{1-\frac{4 m^2_h}{s}}   \right).
\end{aligned}
\end{equation}

Thermally averaged annihilation cross section for the processes $(a)$, $(b)$ and $(c)$ are compute by
\begin{equation}
    \left\langle \sigma v \right\rangle  =\frac{1}{8Tm^2_-m^2_0K_2\left(\frac{m_-}{T} \right)K_2\left(\frac{m_0}{T} \right)}\int_{(m_-+m_0)^2}^\infty \sigma(s)  \left(s-\left(m_-+m_0\right)^2\right)\sqrt{s}K_1\left(\frac{\sqrt{s}}{T} \right) ds,
\end{equation}
and the expression for the process in $(d)$ is
\begin{equation}
    \left\langle \sigma v \right\rangle (d)  =\frac{1}{8T^3m^3_0K_2\left(\frac{m_0}{T} \right)}\int \sigma(s) \left(s-m_0^2\right)^2 e^{-\frac{s-m^2_0}{2m_0 T}}ds,
\end{equation}
where $K_n(x)$ are the modified Bessel functions of the second kind. The total thermally averaged cross section is given by
\begin{equation}
   \begin{aligned}
       \left\langle \sigma v \right\rangle =& \frac{\pi g^4\left( 1+g^2_+/\cos^4 \theta_W  \right)}{3\times 2^9 m^2_0 K_2\left( x \right){}^2} \left[ \left( 4x^2 +3\right) \left(I_{\frac{3}{4}}\left( x \right) - I_{-\frac{3}{4}}\left( x \right)  \right)^2 \right. \\
       & \left. - \left( 4x^2  -1\right) \left(I_{\frac{1}{4}}\left( x \right)- I_{-\frac{1}{4}}\left( x \right)  \right)^2 -\frac{1}{x  \pi^2}\left( 4x^2  -3 \right) K_{\frac{1}{4}}\left( x \right) K_{\frac{3}{4}}\left( x \right)\right] \\
       &-\frac{\pi g^4\left( \cos (2\theta)+2e^2+2 \right)}{3\times 2^{11} m^2_0 x^2  K_2\left( x \right){}^2} \left[ 4x^2 \left( 10x^2   -1 \right) \left(I_{\frac{1}{4}}\left( x \right) - I_{-\frac{1}{4}}\left( x \right)  \right)^2 + \left( 10x^2  +3  \right) \right. \\
       &\left. \left( I_{\frac{1}{4}}\left( x \right){}^2 +8x^2  I_{-\frac{3}{4}}\left( x \right)I_{\frac{3}{4}}\left( x \right)-4x^2  \left(I_{\frac{3}{4}}\left( x \right){}^2+I_{\frac{5}{4}}\left( x \right){}^2   \right) \right) -2x  \Bigg( 3 \left( 10x^2   +1 \right) \right.   \\
       &\left. \left. I_{-\frac{3}{4}}\left( x \right)I_{\frac{1}{4}}\left( x \right) - \left( 10x^2   -3  \right)  \left( I_{\frac{1}{4}}\left( x \right)I_{\frac{3}{4}}\left( x \right)+\frac{\sqrt{2}}{\pi}I_{-\frac{1}{4}}\left( x \right)
       K_{\frac{3}{4}}\left( x \right) \right) \right) \right] \\
       &+\frac{\left|Y_\psi\right|^2 x^4 \cos^2\theta}{2^{13}\pi m^2_\phi m^9_0 K_2\left( x \right)} \left[ \frac{2m_0 m^2_\phi }{x } e^{-\frac{m^2_\phi }{2m^2_0}  x }  \lambda^2 \left[ u^2 \left(  \left( m^2_\phi -m^2_0 \right)^2 +\frac{2m^4_0}{x }\right) +\frac{2v^2}{x }m^4_0  \right] \right.   \\
       & -\frac{6 \lambda_\sigma^2 u^2m^3_0}{x }e^{-\frac{4m^2_\phi-m^2_0 }{2m^2_0}x } \left[\left( m^2_\phi -m^2_0 \right)^2 -\frac{6}{x } m^2_0 m^2_\phi  \right] + e^{-\frac{m^2_\phi-m^2_0 }{2m^2_0}x }m_0 \bigg[ \left( m^2_\phi -m^2_0 \right) \\
       &\lambda^2 Ei\left(-\frac{x }{2}\right)\left[m^2_\phi \left( \left(m^2_\phi -m^2_0  \left(1+\frac{4}{x }  \right) \right)u^2 -\frac{2v^2}{x } \left( m^2_\phi -m^2_0 \right)  \right) \right] \\
       & -9\lambda^2_\sigma u^2 \left( m^2_\phi -m^2_0 \right) \left(m^4_\phi - m^2_\phi m^2_0 -\frac{4}{x } m^4_0\right)Ei\left(-\frac{3m^2_\phi x }{2m^2_0}\right) \\
      & \left. +\frac{2}{x }e^{\frac{m^2_\phi }{2m^2_0}x } m^6_0 \left[ 18\lambda^2_\sigma u^2 Ei\left(-\frac{2m^2_\phi x }{m^2_0}\right)  +\lambda^2 v^2 Ei\left(-\frac{m^2_\phi+m^2_0 }{m^2_0}x  \right) \right] \right],
   \end{aligned}
\end{equation}
where $x = m_0/T$, $I_{\alpha}\left(x\right)$ are the modified Bessel functions of the first kind, and $Ei(x)=\int_{-\infty}^x dt \, (e^{t} /t)$ is the exponential integral function. During this computation, we neglect the SM particle masses as $\sqrt{s}>m_0 > m_f, m_h, m_W, m_Z$, and take $m_- = m_0$ for simplicity.

%
%

\bibliographystyle{RAN}
\bibliography{biblio}

\end{document}